\documentclass[aps,pra,showpacs,twocolumn]{revtex4}
\usepackage{graphicx}
\usepackage{subcaption}
\usepackage{bm}% bold math
\usepackage{amsmath}% equation formating
\usepackage{esvect}

\providecommand{\abs}[1]{\left\lvert#1\right\rvert}

\providecommand{\bra}[1]{\langle #1 \rvert}
\providecommand{\ket}[1]{\lvert #1 \rangle}

\providecommand{\be}{\begin{equation}}
\providecommand{\ee}{\end{equation}}
\providecommand{\ba}{\begin{eqnarray}}
\providecommand{\ea}{\end{eqnarray}}

\usepackage{mathbbol}

\usepackage{tikz}
\usepackage{amsfonts,amssymb}
\usepackage{dsfont}
\usepackage{physics}
\usepackage{tocvsec2}

\usepackage{appendix}

\begin{document}

\title{The Hong-Ou-Mandel experiment: from photon indistinguishability to continuous variables quantum computing}
 
\author{ N. Fabre $^1$, M. Amanti $^2$,  F. Baboux $^2$, A. Keller $^{2,3}$, S. Ducci $^2$ and P. Milman $^2$}

\affiliation{$^{1}$ Departamento de Óptica, Facultad de Física, Universidad Complutense, 28040 Madrid, Spain}
\affiliation{$^{2}$Université Paris Cité, CNRS, Laboratoire Matériaux et Phénomènes Quantiques, 75013 Paris, France}
\affiliation{$^{3}$Univ. Paris-Sud 11, Institut de Sciences Mol\'eculaires d'Orsay (CNRS), B\^{a}timent 350--Campus d'Orsay, 91405 Orsay Cedex, France}

\begin{abstract}

We extensively discuss the Hong-Ou-Mandel experiment taking an original phase-space-based perspective. For this, we analyze time and frequency variables as quantum continuous variables in perfect analogy with position and momentum of massive particles or with the electromagnetic field's quadratures. We discuss how this experiment can be used to directly measure the time-frequency Wigner function and implement logical gates in these variables. We also briefly discuss the quantum/classical aspects of this experiment providing a general expression for intensity correlations that explicitize the differences between a classical Hong-Ou-Mandel like dip and a quantum one. Throughout the manuscript we will often focus and refer to a particular system based on AlGaAs waveguides emitting photon pairs via spontaneous parametric down conversion, but our results can be extended to other analogous experimental systems and to different degrees of freedom.   

\end{abstract}
\pacs{}
\vskip2pc 
 
\maketitle

\section{Introduction}

One of the greatest challenges when describing a scientific revolution is to structure, from the future, a linear and coherent narrative of how ideas emerged, and how they were discussed,  combined, proved and disproved so as to  finally converge into a theory. This difficulty is particularly pronounced in what concerns quantum physics. Not only there is no complete consensus about its interpretation, but also it's not clear wether it is indeed possible to have a single only interpretation of it. In addition, and for some people, most importantly, many of the existing interpretations require completely abandoning some principles of physics which are known as ``classical". In spite of all that, we keep on doing research, predicting and confirming physical phenomena that would not have been imaginable if we hadn't agreed at some point that controversial as it is, using whatever interpretation we chose, quantum physics is the most accurate theory to describe and predict physical phenomena in a given scale. 

Nevertheless, and as a consequence of this debate, a recurrent question arises when observing and studying quantum phenomena: what is so quantum about all that ? Is this really quantum ?  Indeed, it's a hard to answer question, that becomes even harder when followed by its twin sibling: ``what's the classical counterpart of  the situation I am studying? ", ``what should I compare my results to ?" ``Is it possible to compare quantum and classical behavior ?".  

It's clear that quantum computing and quantum information science brought some light into these problems, by defining quantum advantage, or supremacy, as the power of some quantum algorithms and protocols to move problems from one complexity class to another. This is an objective method to separate quantum from non-quantum phenomena. Putting all this into work and using quantum mechanics to solve practical problems by designing quantum machines is what A. Aspect, termed the second quantum revolution.

From the physics and quantum optics point of view, the quantum/classical frontier and questions about how to define it and place it seems less explicit, but we cannot avoid plunging into them when we thumb through Mr. A. Aspect's lectures, in particular \cite{Aspect:19}, seeking for inspiration for writing the introduction of a paper in his honor. What does he mean when he says something is quantum or classical ? Does it make sense ? 

Starting from these questions, we concentrate on some interesting aspects of one of the experiments Mr. A. Aspect uses to illustrate the differences between quantum and classical physics in his lectures, the Hong-Ou-Mandel (HOM) experiment \cite{HOM}. We'll see how this experiment, that we briefly describe and discuss in Section \ref{HOM}, followed by its classical analog (Section \ref{Classical}) can be associated to the time-frequency phase space (Section \ref{SecWigner}), and finally provide some applications of our results (Sections \ref{Boucher}).

\section{The Hong-Ou-Mandel experiment}\label{HOM}

\begin{figure}[h]
\centering
\includegraphics[width=\columnwidth]{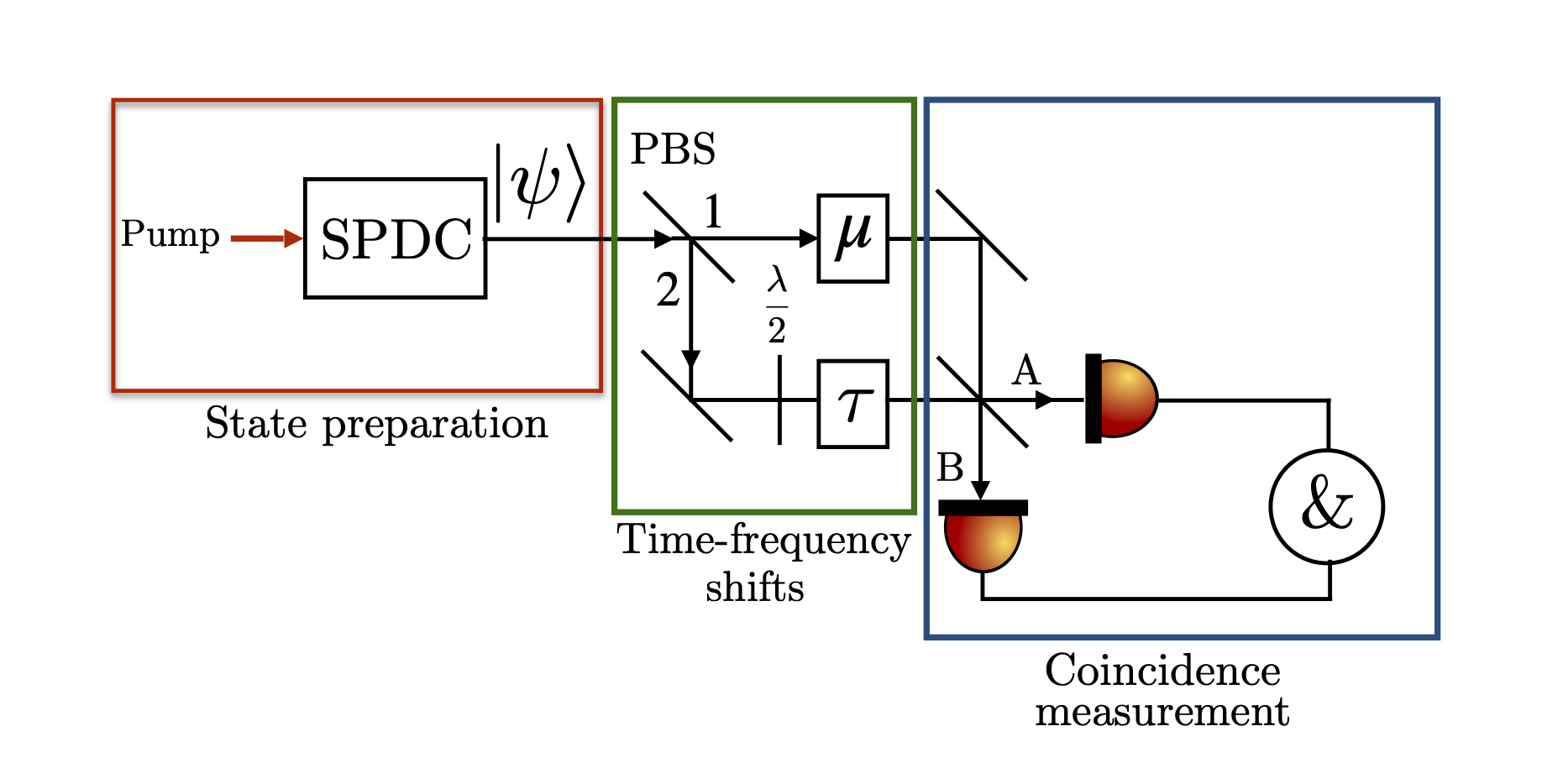}
\caption{Modified HOM experiment including a time delay $\tau$ in one input arm and frequency displacements of $\mu$ in the other input arms. As described in Sections \ref{Boucher} and \ref{SecGKP}, the experiment can also be seen as composed of a preparation step (for state engineering and for an alternative to implement frequency displacements for quantum state measurement, red box), a manipulation step (that can be used to implement logical gates, as described in \ref{SecGKP} or for quantum state measurement as well, greed box) and a detection step based on coincidence measurements (blue box).  }
\label{figHOM}
\end{figure}

Quantum phenomena are usually, if not always, associated to interference effects. The Hong-Ou-Mandel experiment, depicted in Fig. \ref{figHOM} is an example of a clever way to use an interferometer to probe photon indistinguishability using a consequence of the bosonic commutation relations. Indeed the experiment shows that indistinguishable photons (even generated by independent sources with no a priori phase relation) interfere in such way that they both exit the same output port of the interferometer. We call this behavior ``bunching" and its experimental signature is the absence of coincidence detections (the famous HOM dip).  To understand this behavior we consider two photons in different spatial modes, each one of them associated to one input arm of the interferometer. We define the following notation: $\hat a_i^{\dagger}(\omega_i)\ket{0}=\ket{1_{\omega_i}}=\ket{\omega_i}$, so that the corresponding state can be described in the general form  

\begin{equation}\label{state}
\ket{\psi}=\int \int f(\omega_1,\omega_2)\ket{\omega_1, \omega_2}{\rm d}\omega_1 {\rm d}\omega_2,
\end{equation}
where the indices $i=1,2$ denote different spatial modes and $\omega$ is an arbitrary photonic degree of freedom, as the transverse position and momentum or frequency. In the present contribution, we consider pure states for simplifying reasons (even though general expressions are given in Section \ref{Classical}) and we focus on the frequency degree of freedom, since we will take as case study an experimental system consisting of semiconductor waveguides emitting frequency and/or polarization correlated photon pairs by spontaneous parametric down conversion
 \cite{Orieux:11, Photoniques, doi:10.1080/09500340.2014.1000412}. A more detailed description of the experimental device can be found in Section \ref{Boucher}.

In the particular case of photons generated from independent sources we have that $ f(\omega_1,\omega_2)=g(\omega_1)g(\omega_2)e^{i\phi}$ in Eq. (\ref{state}), which means that the spectral amplitude of each photon is the same up to an arbitrary and random phase factor (a consequence from the fact that photons are independent).This is the original context of the first HOM experiment \cite{HOM}. 

The two photons are sent into the two input arms of an interferometer, as shown in Fig. \ref{figHOM}. After the $50/50$ beam-splitter, the input spatial modes are combined in the following way: $\hat a_A(\omega) = (\hat a_1(\omega)+\hat a_2(\omega))/\sqrt{2}$ and $\hat a_B(\omega) = (\hat a_1(\omega)-\hat a_2(\omega))/\sqrt{2}$, where $A$ and $B$ are the output spatial modes where photons are detected in coincidence after leaving the beam-splitter. Thus, the state described by Eq. (\ref{state}) is transformed into

\begin{eqnarray}\label{state2}
&&\ket{\psi}=\\
&&\frac{1}{2}\int \int g(\omega_1)g(\omega_2)e^{i\phi} (\hat a_A^{\dagger}(\omega_1)\hat a_B^{\dagger}(\omega_2)-\hat a_A^{\dagger}(\omega_2)\hat a_B^{\dagger}(\omega_1) +\nonumber \\
&&\hat a_A^{\dagger}(\omega_1)\hat a_A^{\dagger}(\omega_2)-\hat a_B^{\dagger}(\omega_1)\hat a_B^{\dagger}(\omega_2))\ket{0}{\rm d}\omega_1 {\rm d}\omega_2.\nonumber
\end{eqnarray}
Two single photon detectors are used to measure temporal correlations between the photon exiting the two output ports.
We now express the coincidence detection probability. The terms  in the third line of Eq. (\ref{state2}) correspond to photon bunching and do not lead to coincidence detections, since both photons leave the interferometer in the same spatial mode. They are thus ignored from now on and we focus on the two first terms in the r.h.s. of Eq. (\ref{state2}). The coincidence detection probability ${\cal C}$ between photons in modes $A$ and $B$ is given by the square of the absolute value of this term. We have thus: 

\begin{equation}\label{Coincidence}
{\cal C} = \frac{1}{2}- \frac{1}{2}\int \int |g(\omega_1)|^2 |g(\omega_2)|^2 {\rm d}\omega_1 {\rm d}\omega_2,
\end{equation}
and using the normalization condition $\int \int |g(\omega_1)|^2 |g(\omega_2)|^2 {\rm d}\omega_1 {\rm d}\omega_2 =1$, we obtain that ${\cal C}=0$ irrespectively of the phase $\phi$. Consequently, we can conclude that single indistinguishable photons generated by independent sources sent in a balanced interferometer - {\it i.e.}, where both photons go through the same optical path but in different spatial modes - always bunch and exit the interferometer from the same port.

This important result can be made even more interesting if we add some spice to it, in the form of a controllable optical path difference between the input photons. In this case, the initial state right before impinging the beam-splitter in Fig. (\ref{figHOM}) is characterized by the wave-function $ f(\omega_1,\omega_2)=g(\omega_1)g(\omega_2)e^{i\omega_1 \tau}$, where we dropped the phase $\phi$, since it doesn't play a role in the coincidence detection, and added a time and frequency-dependent phase that comes from the path difference between the two arms. It's easy to verify that in this case

\begin{eqnarray}\label{Coincidence2}
&&{\cal C}(\tau) = \frac{1}{2}- \frac{1}{2}\int \int |g(\omega_1)|^2 |g(\omega_2)|^2 e^{i(\omega_1-\omega_2)\tau}{\rm d}\omega_1 {\rm d}\omega_2  \nonumber \\
&&= \frac{1}{2}\left ( 1- |\int \tilde g(t) \tilde g^*(t-\tau){\rm d}t|^2\right )
\end{eqnarray}
where $\tilde g(t)$ is the  Fourier transform of $g(\omega)$ at point $t$ . Eq. (\ref{Coincidence2}) is in general different from zero. If the overlap between the Fourier transforms calculated at the two points appearing in Eq. (\ref{Coincidence2}) is different from one, we say that the two incoming photons are not completely indistinguishable, and in the limit of complete distinguishability, ${\cal C}(\tau) =1/2$. This result is the same one would expect to find if photons were billiard balls arriving into a path bifurcation in pairs and are randomly sent throught one output path or the other,  $A$ or $B$. This is why we refer to this situation as the classical one. 

In spite of having been initially thought to probe the indistinguishability of photons coming from independent sources, the first experimental demonstrations of the Hong-Ou-Mandel effect used photon pairs generated by spontaneous down conversion (SPDC) processes \cite{HOM}. The reason for that is the experimental difficulty at that time to built independent single photon sources, a difficulty overcame in \cite{PhysRevLett.96.240502}.  Nevertheless, as we will see throughout the present contribution, the HOM effect presents many other facets than the one it was initially designed for. Its applications are numerous, and for presenting some of them it's important to provide, for the general case of (\ref{state}), the coincidence probability. Using the same manipulations to obtain (\ref{Coincidence}) we have that 

\begin{equation}\label{Coincidence3}
{\cal C}(\tau) = \frac{1}{2}- \frac{1}{2}\int \int f(\omega_1,\omega_2) f^*(\omega_2, \omega_1) e^{i(\omega_1-\omega_2)\tau}{\rm d}\omega_1 {\rm d}\omega_2,
\end{equation}
which is the central equation of the present manuscript. Simple as it seems to be, Eq. (\ref{Coincidence3}) can provide information about entanglement \cite{eckstein_broadband_2008}, time-frequency phase space representation \cite{douce_direct_2013}, and be used to characterize resources for quantum simulation \cite{francesconi_engineering_2020, Francesconi21}, metrology \cite{chen_hong-ou-mandel_2019} and quantum error correction \cite{PhysRevA.102.012607}.  

\subsection{On the ``quantumness" of the HOM experiment}\label{Classical}

Before moving to the central point of the present contribution we'll discuss a little further the interpretation of the HOM experiment as probing quantum properties of light. In \cite{PhysRevA.100.013839}, an experiment involving intensity correlations between classical fields with a fixed, but random phase reference showed that it is indeed possible to reproduce the HOM dip with close to $100 \%$ visibility. While the theoretical model presented in \cite{PhysRevA.100.013839} uses a classical description of light, we provide here a theoretical description using intensity correlations between coherent states which are sent at the two input ports of a HOM-like interferometer and where intensity correlations are calculated between the signals recorded at the two output ports, as in \cite{PhysRevA.100.013839}. 

Using the same spatial modes labelling  given in the introduction, we have that,  after the introduction of a time delay $t$ in one arm (say, arm 2) of the HOM interferometer and recombination in the beam-splitter, the normalized intensity correlation between modes $A$ and $B$ can be written as 
\begin{equation}\label{NIC}
{\cal C}(t)= \frac{\langle \hat N_A \hat N_B \rangle}{\langle \hat N_A \rangle \langle \hat N_B \rangle},
\end{equation}
where
\begin{equation}
\hat N_i=\int \hat a_i^{\dagger}(\omega)\hat a_i(\omega){\rm d}\omega,
\end{equation}
with $i=A,B$, and where the average is taken both over classical parameters (as the phase $\phi)$ and on the quantum state. Eq. (\ref{NIC}) can be rewritten using :
\begin{eqnarray}\label{NICAB}
&&\hat N_{A(B)}= \frac{1}{2}\int \left (\hat a_1^{\dagger}(\omega)e^{-i\omega t} \pm \hat a_2^{\dagger}(\omega)\right ) \times \nonumber \\
&&\left (\hat a_1(\omega)e^{i\omega t}\pm \hat a_2(\omega)\right ) {\rm d}\omega \nonumber \\
&&= \frac{1}{2}\left [ \hat N_1 + \hat N_2 \pm (\hat {\cal I}_{1,2}(t) +\hat {\cal I}_{1,2}^{\dagger}(t))\right ],
\end{eqnarray}
with $\hat {\cal I}_{1,2}(t)= \int \hat a_1^{\dagger}(\omega)\hat a_2(\omega)e^{i\omega t} {\rm d}\omega$ and $\hat N_{1(2)}=\int \hat a_{1(2)}^{\dagger}(\omega)\hat a_{1(2)}(\omega){\rm d}\omega$. Finally, ${\cal C}(t)$ can be written as: 
\begin{equation}\label{NICFinal}
{\cal C}(t)= \frac{\langle \hat N ^2 \rangle - \langle (\hat {\cal I}_{1,2}(t) +\hat {\cal I}_{1,2}^{\dagger}(t))^2 \rangle}{(\langle \hat N  \rangle^2 -\langle \hat {\cal I}_{1,2}(t) +\hat {\cal I}_{1,2}^{\dagger}(t) \rangle^2)},
\end{equation}
where $\hat N=\hat N_1 + \hat N_2$ is related to the total field intensity. Expression (\ref{NICAB}) is general and can be used also for non-pure states \cite{PhysRevLett.126.063602}, and a similar expression was obtained in \cite{MJWoolleyetal2013} expressed as temporal correlations for microwave fields. As a matter of fact, it is easy to verify that this expression reduces to the coincidence probability if we consider as input state a biphoton by noticing that in this case
\begin{eqnarray}\label{IAB}
&&\hat {\cal I}_{1,2}(t)\hat {\cal I}_{1,2}^{\dagger}(t)=  \hat N_1  + \\
&& \int \int \hat a_1^{\dagger}(\omega_A)\hat a_2^{\dagger}(\omega_B)\hat a_1(\omega_B)\hat a_2(\omega_A) e^{i(\omega_A-\omega_B)t}{\rm d} \omega_A {\rm d} \omega_B,\nonumber
\end{eqnarray}
and 
\begin{eqnarray}\label{IBA}
&&\hat {\cal I}_{1,2}^{\dagger}(t)\hat {\cal I}_{1,2}(t)= \hat N_2  + \\
&&  \int \int \hat a_1^{\dagger}(\omega_B)\hat a_2^{\dagger}(\omega_A)\hat a_1(\omega_A)\hat a_2(\omega_B) e^{i(\omega_B-\omega_A)t}{\rm d} \omega_A {\rm d} \omega_B, \nonumber
\end{eqnarray}
and $\langle \hat {\cal I}_{1,2}^{\dagger}(t) \rangle = \langle \hat {\cal I}_{1,2}(t) \rangle = \langle\hat {\cal I}_{1,2}^{\dagger}(t)^2\rangle = \langle\hat  {\cal I}_{1,2}(t)^2 \rangle = 0$.

\subsection{Application to coherent states}

We now compute (\ref{NICAB}) in the case where $\ket{\psi}=\ket{\alpha}_1\ket{\beta}_2$, with $\beta=|\alpha | e^{i\phi}$ and $\alpha = |\alpha|$, in analogy to the calculations in \cite{PhysRevA.100.013839} done using classical fields. We have that

\begin{eqnarray}\label{coherent}
&&{\cal C}(t)=\frac{1}{\langle \hat N \rangle^2_{\gamma} - 4({\rm Re}\left [ \int  |\alpha(\omega)|^2 e^{i\phi}e^{i\omega t}{\rm d}\omega \right ])^2} \times \nonumber \\
&&  ( \langle \hat N \rangle^2_{\gamma} - 2 {\rm Re}\left [ (\int |\alpha(\omega)|^2 e^{i\phi}e^{i\omega t}{\rm d}\omega)^2\right ] -  \\
&&2 {\rm Re}\left [ \int \int |\alpha(\omega_A)|^2 |\alpha(\omega_B)|^2 e^{i(\omega_A-\omega_B) t}{\rm d}\omega_A{\rm d}\omega_B\right ] )\nonumber , 
\end{eqnarray}
where $\langle \hat N\rangle_{\gamma}=2\int |\alpha(\omega)|^2{\rm d}\omega$. Note that we haven't averaged over $\phi$ yet: in order to do so, let's take a look at expression (\ref{coherent}) for $t=0$. In this case, we have that ${\cal C}(0)= \frac{\langle \sin^2 {\phi} \rangle_{\phi}}{1-\langle \cos{\phi}\rangle_{\phi}^2}$. From this expression, we see that if the phase $\phi$ is uniformly distributed in the continuous interval ranging from $0$ to $2\pi$, we have ${\cal C}(0)=1/2$ with a visibility of $1/2$. If the phase can take only two values with equal probability and $\phi \in \{0, \pi\}$, we have that ${\cal C}(0)=0$ with a theoretical visibility of $1$, which reproduces the two-photon coincidence case in the classical regime. Finally, if the two possible and equally probable phases are such that $\phi \in \{\pi/2,3\pi/2\}$, ${\cal C}(0)=1$ and the visibility is zero. These results are in accordance with the ones in \cite{PhysRevA.100.013839}, and the authors use them to argue that observing a near $100 \%$ visibility dip in intensity correlations is not necessarily a quantum signature. Accordingly, the authors suggest that by adding a wave-particle duality-like experiment in the output of the HOM interferometer one could then obtain observable differences between quantum and classical properties, which is of course also true. 

The conclusion that it is possible to mimic the HOM dip with classical fields should be put into perpective by inspecting Eqs. (\ref{coherent}) and (\ref{NICFinal}). We can see that the average zero correlation effect is caused by the $\phi$ dependent term, which is a first order interference term. One can thus easily identify and isolate it, and argue that quantum effects are associated to second-order correlation terms only \cite{Bouchard_2020}. By keeping only these terms, it is clear that the HOM dip cannot be reproduced with classical fields. Another way to argue in this sense is by noticing that the $\phi$ dependent term presents rapid oscillations that are averaged to zero depending on the detection sensibility to fluctuations in a given frequency range. This is precisely the situation of the Hanbury Brown and Twiss experiment, for instance, where only purely second order correlation terms remain and such first order phase-dependent effects do not play a role. 

From our point of view, it is interesting to notice that the spectral intensity overlap between classical fields in Eq. (\ref{coherent}) doesn't lead to $100 \%$ visibility of the intensity correlations while being always $\leq 0$, so that in the absence of the first order interference term intensity correlations are always $1/2 \leq {\cal C}(t) \leq 1/4$, with visibility $1/2$. 
 
As a conclusion, even though \cite{PhysRevA.100.013839} presents an interesting classical situation analogous to the HOM dip, it is not clear in our view how to compare both situations and wether this classical result challenges the usual interpretation of the HOM experiment.

\section{The Wigner function}\label{SecWigner}

We now turn back once again to the biphoton situation. Those working in optics, and used to interference effects, are familiar to expressions as (\ref{Coincidence3}), where two amplitudes overlap. The relation between the two overlapping amplitudes in (\ref{Coincidence3}) is such that it may ring a bell in those who are also familiar with the phase space representations of  quantum mechanics. As a matter of fact, Eq. (\ref{Coincidence3}) is pretty similar to the Wigner function of a massive particle, or the Wigner function associated to the quadrature state of a single mode electromagnetic field, which is given by: 
\begin{equation}\label{Wigner}
W(x,p)= \int \bra{x-\frac{q}{2}}\hat \rho \ket{x+\frac{q}{2}}e^{ips}{\rm dq} . 
\end{equation}
For a pure state, Eq. (\ref{Wigner}) reduces to

\begin{equation}\label{Wigner2}
W(x,p)=\int \psi^*(x-\frac{q}{2})\psi(x+\frac{q}{2})e^{ipq}{\rm d q} ,
\end{equation}
where $x$ and $p$ are the position and momentum variables or, equivalently, two orthogonal quadratures of the electromagnetic field, and $\psi(x)$ is the wave-function in, say, the position basis. We see that Eqs. (\ref{Wigner2}) and (\ref{Coincidence3}) are indeed very similar but still, not identical. 

Indeed, when comparing (\ref{Wigner2}) to (\ref{Coincidence3}) many differences  appear: in first place, it seems that we have the equivalent to a two particle system in Eq. (\ref{Coincidence3}), since we have a two-variable integral (in $\omega_1$ and $\omega_2$ variables). Also, there is no displacement in the frequency variables, as there is in the position one in (\ref{Wigner2}). Finally, (\ref{Wigner2}) describes the quantum state of a particle or of the electromagnetic field, but what about (\ref{Coincidence3}) ? We will handle these issues one by one to conclude that, indeed,  (\ref{Coincidence3}) and (\ref{Wigner2}) are pretty much the same, or at least provide the same type of information about the quantum state of a system in the case one limits the discussion to the case where each mode is occupied by one photon only. 

To start with, and for pedagogical reasons, we perform a change of variables in (\ref{Coincidence3}) such that: $\omega_{\pm}=(\omega_1 \pm \omega_2)/2$. In this case,  $f(\omega_1,\omega_2)=f'(\omega_+,\omega_-)$ and $f(\omega_2,\omega_1)=f'(\omega_+,-\omega_-)$. Then, following \cite{douce_direct_2013}, we define $g( \omega_-)g^*(- \omega_-)=\int f'(\omega_+, \omega_-)f'^*(\omega_+, -\omega_-) {\rm d}\omega_+$. In the case where $ f'(\omega_+, \omega_-)$ is a separable function in the $\omega_{\pm}$ variables, $g( \omega_-)$ is the wavefunction associated to the $\omega_-$ variable and provides all the information about this collective mode. In the general case, we're considering a marginal of the wave-function. In any case, using the previously introduced definition, we have that: 

\begin{equation}\label{Coincidence4}
{\cal C} (\tau)= \frac{1}{2}- \frac{1}{2}{\rm Re}[\int  g(\omega_-)g^*(-\omega_-) e^{i\omega_-\tau}{\rm d}\omega_- ],
\end{equation}
where we have transformed the double integral of Eq. (\ref{Coincidence3}) into a single one, in the form of (\ref{Wigner2}). 

Now we deal with the issue of the equivalent to the position displacement appearing in Eq. (\ref{Wigner2}), which in Eq. (\ref{Coincidence4}), a displacement in variable $\omega_-$ (while momentum displacement in (\ref{Wigner2}) is equivalent to the $\omega_-$ dependent time displacement appearing in (\ref{Coincidence4})). It's clear that transformations of the type $\pm \omega_- \rightarrow \mu \pm \omega_-$ would transform (\ref{Coincidence4}) into the exact mathematical analog of  (\ref{Wigner2}) with the replacements $\tau \rightarrow p$, $\omega_- \rightarrow 2q$ and $x \rightarrow \mu$. By analyzing the interferometer in Fig. \ref{figHOM} we see that a frequency displacement of $\mu$ in frequency in one of the interferometer's arms would modify the photonic state and its  wave-function in such a way that the coincidence detection at the outputs of the interferometer would give 

\begin{equation}\label{Coincidence5}
{\cal C}(\mu, \tau) = \frac{1}{2}- \frac{1}{2}{\rm Re}[\int  g(\mu +\omega_-)g^*(\mu -\omega_-) e^{i\omega_-\tau}{\rm d}\omega_-]. 
\end{equation}

We leave the discussion about possible ways to experimentally implement such displacements to Section \ref{Boucher}. The integral in Eq. (\ref{Coincidence5}) has now the same form as the one in (\ref{Wigner2}). So, what's the meaning of all that ? Is this really a Wigner function, as  (\ref{Wigner2}) ? And if it's the case, what type of information can it provide about the state, or at least about the part of the state associated to variable $\omega_-$ ? These issues were addressed in \cite{douce_direct_2013, boucher_toolbox_2015, fabre_producing_2020}, leading to the conclusion that indeed, the coincidence detection of the output of a HOM interferometer can be expressed as: 

\begin{equation}\label{Coincidence6}
{\cal C} (\mu, \tau)= \frac{1}{2}-\frac{1}{2}W_-(\mu,\tau), 
\end{equation}

where $W_-(\mu,\tau) = \int  g(\mu +\omega_-)g^*(\mu -\omega_-) e^{2i\omega_-\tau}{\rm d}\omega_-$ is the {\it chronocyclic Wigner function}  associated to the variable $\omega_-$ of the photon pair. In the case of a biphoton wavefunction which is separable in the variables $\omega_-$ and $\omega_+$, this information completely characterizes the quantum state of the biphoton associated to the variable $\omega_-$. In particular, we can notice that the usual shape of the HOM dip is nothing but a cut in the phase space of varying $\tau$ and at $\mu = 0$ of a Gaussian spectral function. 

An important point to address is the fact that we're dealing here with the quantum description of the photon pair detected in coincidence, so the Wigner function describes the photonic quantum state in frequency.  The quantum statistics of the field it refers to give to $W_-(\mu, \tau)$ specific properties having a quantum interpretation. For instance, in \cite{eckstein_broadband_2008}, it was shown that if ${\cal C} (0,\tau) > 1/2$, then the photon pair is frequency entangled. We can show that having ${\cal C} (\mu,\tau) > 1/2$ is also an entanglement witness, and this corresponds precisely to saying that the existence of negative points of $W_-(\mu,\tau)$ is a proof that the photon pair is frequency entangled. 

Interpreting the integral in (\ref{Coincidence5}) as a Wigner function contributes to providing a physical picture of frequency and time of single photons as quantum continuous variables. Even though it's  known that such degrees of freedom are continuous, in most protocols and applications they are either discretized \cite{Qudits, PhysRevA.66.062308, Lu:19} or distributed in basis of discrete modes \cite{lamata_dealing_2005, PhysRevLett.84.5304} so that their genuine continuous character is not fully exploited. As shown in \cite{PhysRevA.102.012607}, the single photon time and frequency variables can also be used for quantum information, computing and communication protocols analogously to  ``usual" quantum continuous variables in quantum optics - as the vibrational state of trapped ions or the quadratures of the electromagnetic field. This fact opens the perspective of broadening even more the applications of single photon-based quantum information related protocols, since frequency and time variables are shown to be the continuous version of polarization degrees of freedom, for instance. 

As a concluding remark for this part, we have seen that the coincidence probability in a HOM experiment is a direct measurement of the Wigner function associated to the $\omega_-$ variable of a photon pair. This may seem restrictive, but it is only a consequence of the choice of the experimental system and its relevant degrees of freedom. We have shown in \cite{douce_direct_2013} that the HOM provides a way to characterize other photonic degrees of freedom as well, as for instance the Wigner function associated to the transverse position and momentum coordinates. For these degrees of freedom, it is easier to perform spatial rotations of the field's profile so that one can directly access different combinations of variables, as for instance the one associated not only to the difference but also to the sum of the transverse coordinates of the photons. 

It is also possible to adapt the ideas proposed in \cite{douce_direct_2013} so as to access the full Wigner function of a single photon. In this case, instead of seeing the photon pair produced by a SPDC process as produced by the physical device, we can interpret it as the product of a series of logical gates that are applied to a separable state. For this, it was shown in \cite{fabre_spectral_2022, PhysRevA.105.052429},  that the frequency beam-splitter operation defined as 

\begin{equation}
\hat{U}\ket{\omega_{1},\omega_{2}}=\ket{\omega_{+},\omega_{-}}.
\end{equation}
plays the role of the entangling interaction that naturally occurs in the SPDC process, which are physically associated to the phase matching condition and energy conversation. Thus, the produced entangled state that outputs a SPDC device can formally be obtained by starting from two initially separable single photons described by the wavefunction $\ket{\psi}=\iint d\omega_{1} d\omega_{2} f(\omega_{1})g(\omega_{2}) \ket{\omega_{1},\omega_{2}}$ which are then entangled by a frequency beam-splitter operation. 

As a result of this interpretation we can interpret the function $f$ (resp. $g$) of the separable initial photon pair as being the same as  the function modeling the energy conservation (resp. the phase-matching conservation) of a SPDC process. Thus, the coincidence measurement  corresponds to measuring  the spectral function of an initial single photon that has been entangled (changed variables) by the device. In Ref. \cite{fabre_spectral_2022}, we also proposed a direct way for implementing such an operation with non-linear crystals in cascade. However, more efficient ways could be investigated within light-matter interaction using split-ring resonators for instance.

Finally, if a separable two-photon state enters into the  HOM interferometer, the coincidence probability corresponds to the spectrogram:
\begin{equation}
S(\mu,\tau)=\frac{1}{2}(1-\abs{\int f^{*}(\omega_{1})g(\omega_{1}-\mu)e^{ i\omega_{1}\tau}{\rm d}\omega_1}^{2}).
\end{equation}
The spetrogram is the absolute value of the windowed Fourier transform.  If one spectral function is known, let us say $g$ which can be considered as a window function, a phase of retrieval algorithm allows reconstructing the amplitude and phase of the spectral function $f$.

As a conclusion, the HOM interferometer provides various ways for measuring the amplitude and phase of the spectral function of single and two-photon states.

\section{Quantum state manipulation and engineering}\label{Boucher}

We have seen that interpreting the HOM experiment as a direct measurement of the Wigner function of the biphoton associated to the variable $\omega_-$ requires being capable of implementing displacements in the time-frequency phase space. While time displacement are in practice easy to implement in quantum optics by temporal delay lines, frequency displacement require more involved experimental techniques at the single photon level, some of them based on non-linear optic. Nevertheless, frequency displacements are necessary not only to have access to the whole chronocyclic phase phase space and characterize the single photon state, but also for quantum state engineering. A possibility to implement frequency shifts is by using electro-optic modulators which are already available, working at the single photon level. Such devices are used to perform the spectral tomography of single photons \cite{davis_experimental_2018} and the direct signature of such frequency shifts was observed through the decrease of the visibility of the dip in HOM interferometry in \cite{chen_single-photon_2021}.\\
 
In this section we take a different direction and  provide solutions to implement phase space displacements using techniques as pump engineering and post-selection. 

To start with, we provide the details of a specific experimental system to illustrate our ideas. 

\subsection{An experimental context}

\begin{figure}[h]
\centering
\includegraphics[width=\columnwidth]{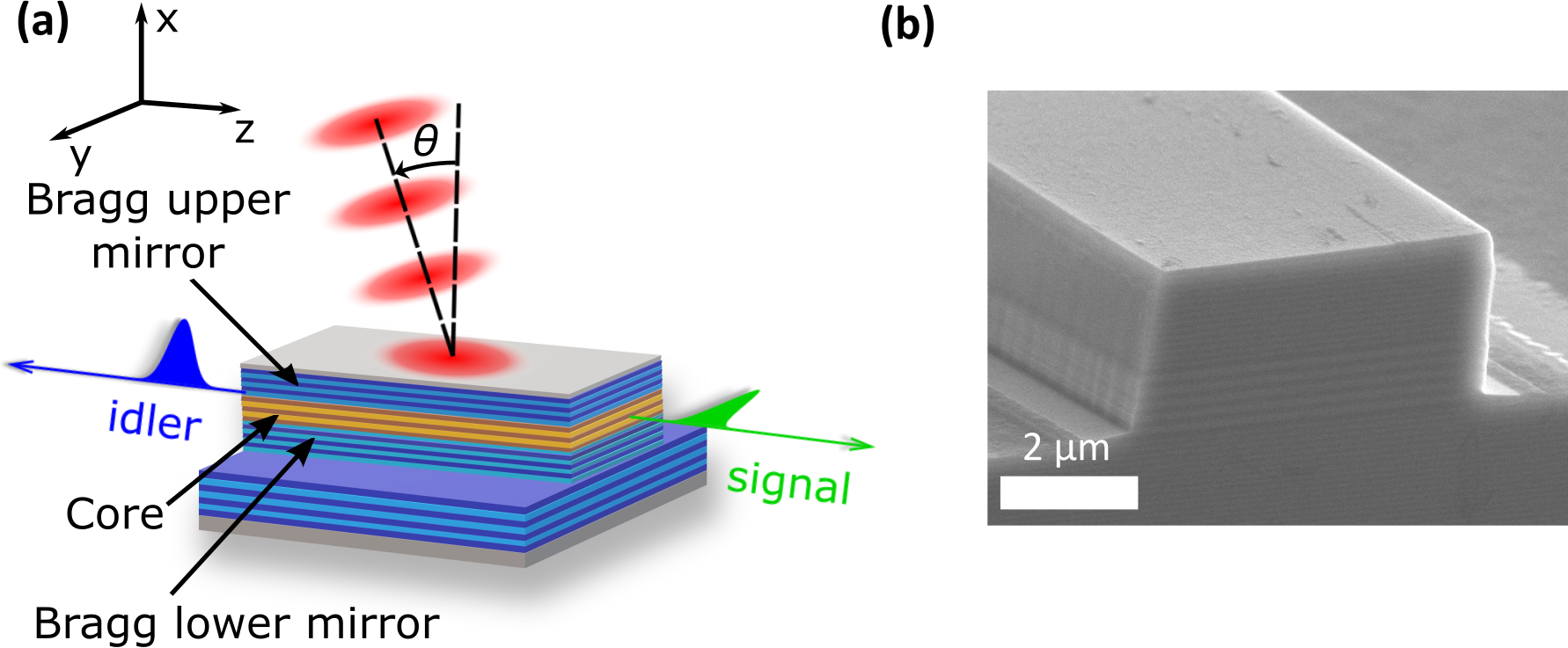}
\caption{
(a) Working principle and (b) SEM image of an AlGaAs ridge microcavity generating frequency-entangled photon pairs by SPDC in a transverse pump geometry.
}
\label{Fig_Source_contra}
\end{figure}

Counterpropagating phase-matching offers a high versatility to engineer the spectral wavefunction and the exchange statistics of photon pairs, as probed in the HOM experiment.
A schematics of our semiconductor chip-integrated source of photons pairs is shown in Fig. \ref{Fig_Source_contra}a, and a SEM image is shown in Fig. \ref{Fig_Source_contra}b.
The source is a Bragg ridge microcavity made of a stacking of AlGaAs layers with alternating aluminum concentrations \cite{Orieux11,Orieux13}. It is based on a transverse pump scheme, where a pulsed laser beam impinging on top of the waveguide (with an incidence angle $\theta$) generates pairs of counterpropagating and orthogonally polarized photons (signal and idler) through SPDC. 
Along the epitaxial direction momentum conservation is implemented through a quasiphase-matching structure of AlGaAs layers (shown in orange in Fig. \ref{Fig_Source_contra}a) with alternating high and low values of the effective nonlinear coefficient.
The Bragg mirrors provide both a vertical microcavity to enhance the pump field and a
cladding for the twin-photon modes.

Two nonlinear interactions can occur simultaneously in the device \cite{Orieux13}; we consider here the one that generates a TE-polarized signal photon (propagating along $z>0$, as sketched in Fig. \ref{Fig_Source_contra}a) and a TM-polarized idler photon (propagating along $z<0$).
When the photons are generated
close to degeneracy and for a narrow pump spectrum, the
biphoton state can be written in the following form:
\begin{equation}
	\ket{\psi}= \iint d \omega_s d \omega_i  f_+(\omega_s+\omega_i) f_-(\omega_s-\omega_i) \ket{\omega_s,\omega_i}
\end{equation}
The joint spectral amplitude (JSA) $\phi(\omega_s,\omega_i)= f_+(\omega_+) f_-(\omega_-)$, where $\omega_\pm=\omega_s \pm \omega_i$, gives the probability amplitude to measure a signal photon at frequency $\omega_s$ and an idler photon at frequency $\omega_i$. 
The function $f_+$, reflecting the condition of energy conservation, is given by the spectrum of the pump beam, while the function $f_-$, corresponding to the phase-matching condition, is determined by the spatial properties of the pump beam:
\begin{equation}\label{Eq2}
	f_-(\omega_-)=\int_{-L/2}^{L/2} dz \, \Phi (z) e^{-i (k_{\rm deg}+\omega_-/\bar{v}_{\rm g})z}
\end{equation}
Here, $\Phi  (z)$ is the pump amplitude profile along the waveguide direction, $L$ is length of the waveguide, $\bar{v}_g$ is the harmonic mean of the group velocities of the twin-photon modes and $k_{\rm deg}=\omega_p\text{\rm sin}(\theta_{\rm deg})/c$, with $\omega_p$ the pump central frequency, $c$ the velocity of light and $\theta_{\rm deg}$ the pump incidence angle corresponding to the production of frequency-degenerate photons.

\subsection{Pump beam engineering}

Measurements in quantum physics can be performed in different ways. Usually, experimental measurement apparatus can only measure quantum states in a given specific basis. Thus, a solution to circumvent this and obtain full information about the state is to manipulate the state so that a measurement in a given state means that, in fact, the system was previously in another state which is considered to be the measured one. 

One can criticize such a technique, since the measured state is not the state that was actually produced, and the limit between state production and measurement is not clearly defined. In the present section, we take a loose position on this debate, and propose quantum state measurement techniques based on quantum state engineering.

In the previous subsection we have seen that the JSA of the produced photon pair can be controlled by modifying the pump laser beam properties, as its angle and position of incidence. In \cite{Boucher15}, we have exploited this fact to propose methods not only to engineer exotic and useful quantum states of the photon pair, but also to measure them by implementing displacements in different directions of the phase space.

\begin{figure}
 \begin{center}
\includegraphics[width=0.5\textwidth]{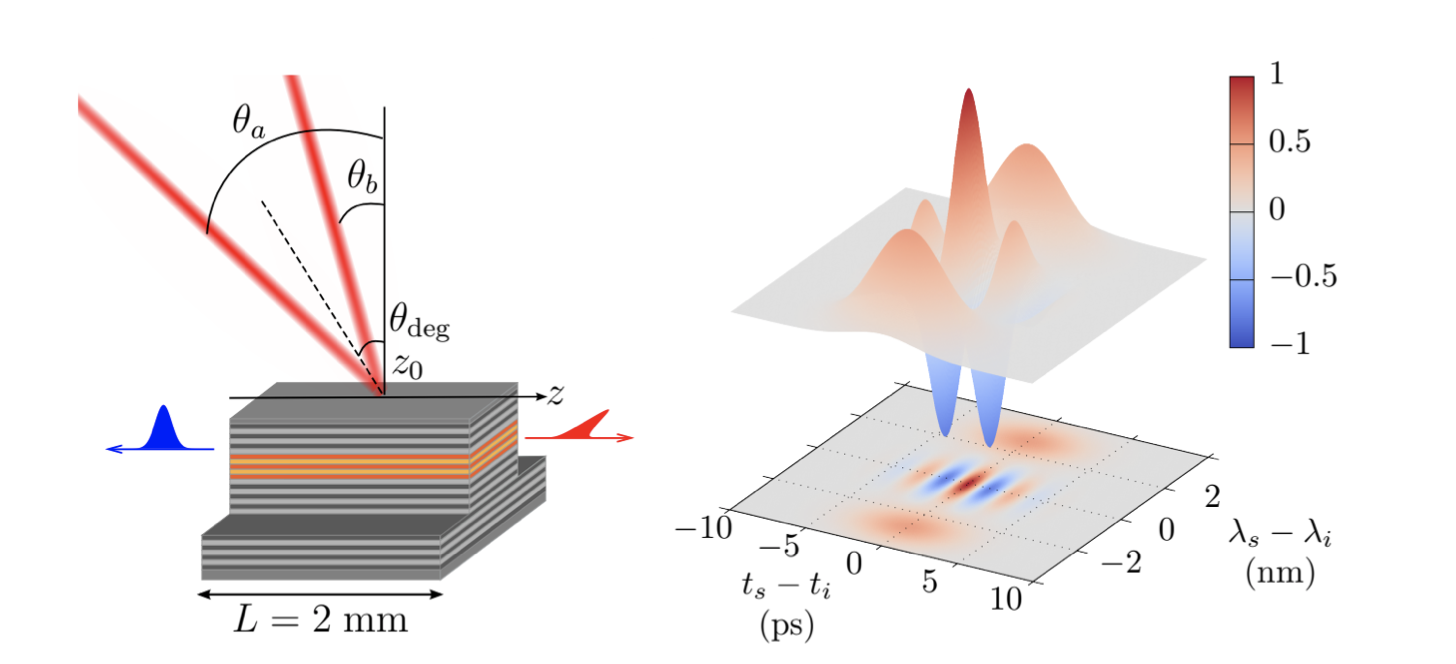}
\caption{\label{Chat} Pump engineering for producing a Schr\"odinger cat-like state in frequency. A divided pump that impinges on the medium at different angles lead to different phase matching conditions which are angle dependent and center the state's frequency distribution at different values. By changing both pump's angles while keeping their angular distance constant one displaces the center of both frequency distributions in the same direction and by the same amount. This is equivalent to performing frequency displacements in phase space, a strategy that can be used both for quantum state engineering and for measuring the Wigner function using the HOM interferometer at different points of the time frequency phase space. }
\end{center}
\end{figure}

Using the results of the previous section, if the dimensions of the waveguide are large with respect to the pump waist, {\it  i.e.} in the limit where $L\rightarrow \infty$, $f_-$ can be approximated as the Fourier Transform of the spatial profile of the pump beam:
\begin{equation}
f_- (\omega_-) \approx \tilde{\varphi}\left(\frac{\omega_-}{\bar{v}_g}\right)
\end{equation}
We start by considering the situation depicted in Fig.~\ref{Chat} where a gaussian pump beam with waist $w_p$ impinges onto the source at an angle $\theta$ and position $z_0$. The field distribution along the $z$ axis is %
$\Phi(z) \propto e^{-{(z-z_0)^2 \cos^2{\theta}}/{w_p^2}} e^{ i (k\sin\theta) z}$ and therefore $f_-$ reads:%

\begin{equation}
f_-(\omega_-)\propto e^{-i\omega_- \tau_0} e^{-\tfrac{\left(\omega_- - \omega_-^{(0)}\right)^2 }{\Delta\omega^2}}
\label{eq:coherentState}
\end{equation}
with $\tau_0 = z_0/\bar{v}_g$, $\Delta\omega = \bar{v}_g\cos{\theta}/\pi w_p$ and $\omega_-^{(0)} = (k\sin\theta-k_{\mathrm{deg}})\bar{v}_g$. 

From a general complex amplitude representation describing a pure state $f_-(\omega_-)$, the Wigner function $W(\tau, \mu)$ at points $\tau, \mu$ of the phase space is given by: 
\begin{equation}\label{eq:wigner}
W(\tau, \mu)=\int_{-\infty}^{\infty}d\omega_-f_-(\mu-\frac{\omega_-}{2})f_-^*(\mu+\frac{\omega_-}{2})e^{i\tau \omega_-}. 
\end{equation}
Using for instance the expression obtained in \eqref{eq:coherentState}, it is easy to see that this corresponds to a gaussian Wigner function centered at point $\tau=\tau_0$, $\mu = \omega_-^{(0)}$. In this situation, shifting the pumping spot $z_0$ is equivalent to realizing displacements in the $\tau$ axis of the phase space while changing the angle of incidence $\theta$ of the pump beam corresponds to shifting the state along the $\mu$ axis. 

More complex states can be obtained by engineering the pump beam. Indeed two identical beams impinging at $z_a$ and $z_b$  generate a superposition of 2 coherent states displaced along the $\tau$ axis, which is a state analogous to a Schr\"odinger cat in position and momentum phase space. The orthogonality between the two Gaussian states is ensured if  $\frac{(z_a - z_b)}{\bar{v}_g} \gg \frac{1}{\Delta\omega}$. %
% as shown in Fig.~\ref{catInTime} for $z_{a,b} = \pm 500\micro\meter$. 
A superposition of two Gaussian packets can also be obtained along axis $\mu$ by using 2 different angles of incidence $\theta_a$ and $\theta_b$ impinging at the same point $z_0$ (see Fig. \ref{Chat}), and this type of state was shown to be a resource for quantum metrology in \cite{chen_hong-ou-mandel_2019, PhysRevA.104.022208}.  % %This situation is depicted in Fig.~\ref{catInFrequency} where $\theta_{a,b} = \theta_{\mathrm{deg}} \pm 1\degree$ 
We can generalize even more these Schr\"odinger cats to arbitrary configuration of pump beams, for instance a compass state, which is a superposition of four coherent states whose utility for metrology has been pointed out elsewhere\cite{PhysRevA.73.023803} and studied in the context of SPDC in \cite{PhysRevA.104.022208}. To obtain it, a set of 4 beams is required: 2 pairs of beams impinging at 2 different points separated by a distance $\Delta z$, each pair consisting of 2 beams symmetrically tilted with respect to the degeneracy angle.

As a conclusion, we can see that pump engineering is a valuable tool not only to design interesting frequency-time entangled state but also to implement displacements in phase space thus enabling quantum state measurement. 

\subsection{Exploiting cavity effects}\label{SecGKP}

In the following, we describe the possibilities of quantum state manipulation and engineering enabled by the the interplay between cavity effects and
temporal delay between photons of a pair. The method can be adapted and applied to a large variety of systems, either bulk or integrated, thus increasing their flexibility and the richness of the generated states.  Here we  illustrate these concepts by taking the example of the biphoton state generation in an AlGaAs waveguide based on a modal phase matching scheme in a collinear geometry.  In this case, the phase-velocity mismatch among the three interacting beams is compensated by a multimode waveguide dispersion engineering by confining the fields using Bragg reflectors \cite{Qudits}.

The phase matching condition depends not only on the properties of the pump, as detailed in the previous section, but also on the properties of the device. Then, another possibility for quantum state engineering is using (and modifying, but this is a more difficult issue)  the latter. Usually SPDC based devices have a phase-matching condition which depends on the temperature, for instance. Using this effect, frequency displacements enabling the direct measurement of the Wigner function $W_-(\mu,\tau)$ were implemented \cite{PhysRevLett.115.193602}. 

Other ways to engineer the  non-linear susceptibility of a non-linear crystal  can be found in \cite{PhysRevA.93.013801} for the single photon and in \cite{https://doi.org/10.48550/arxiv.2204.10079} for the two-mode squeezed regime.

In the following, we describe some interesting properties of the AlGaAs source that can be used for quantum state engineering which are not based on temperature variation effects and turn out to be extremely interesting for quantum state engineering. 

\begin{figure}
\begin{subfigure}{.5\textwidth}
  \centering
  \includegraphics[width=1\linewidth]{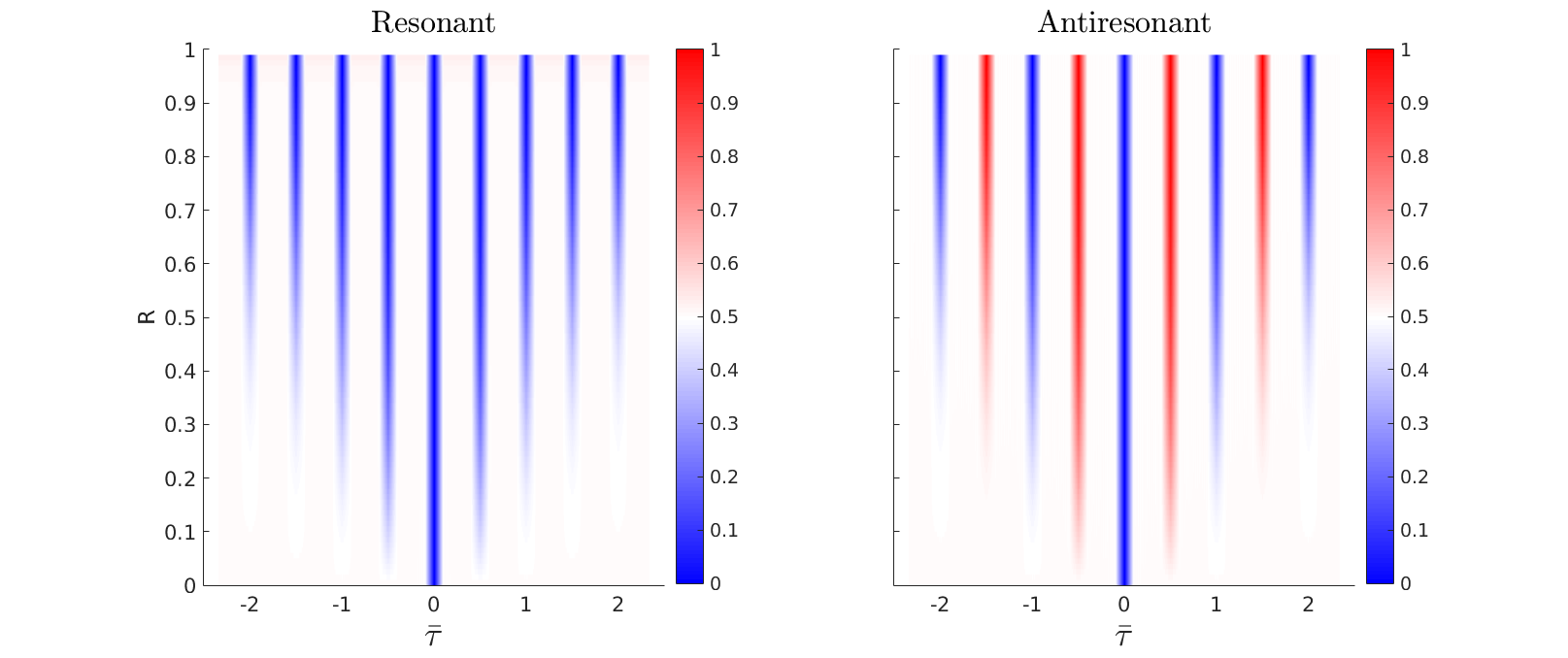}
  \caption{Horizontal axis: time spacing between peaks in units of $\bar \tau/2$. Vertical axis: cavity reflectivity $R$. In the existing experimental device, used in \cite{Qudits}, $R=0.3$. The color code refers to the coincidence probability that oscillates between $0$ and $1$ according to the time delay. $P=1$ for an anti-symmetric state and $P=0$ for a symmetric one. The symmetry of the produced state is related to the resonance or anti-resonance of the pump beam with the cavity frequency and also to the produced states. On the left, the pump beam frequency is resonant with the cavity frequency: peaks are spaced of $\bar \tau/2$, half the time taken for a photon to make a roundtrip in the cavity, and even and odd multiples of $\bar \tau/2$ peaks have the same phase. On the right, the pump beam is anti-resonant with the cavity frequency and peaks with different parity have different phases. The corresponding produced states can be associated to GKP-like states in different basis.}
  \label{fig:sfig1}
\end{subfigure}%

 \vspace{\floatsep} 

\begin{subfigure}{.5\textwidth}
  \centering
  \includegraphics[width=1\linewidth]{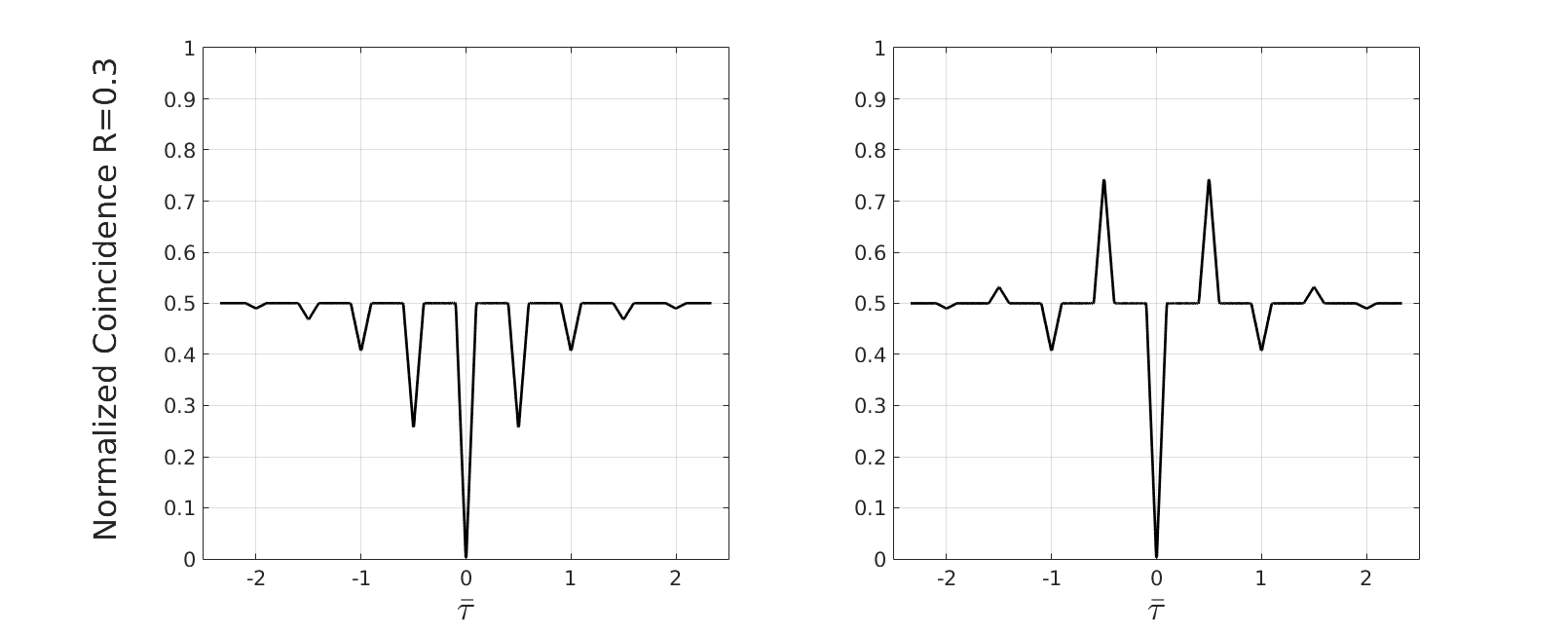}
  \caption{Simulated HOM using as the cavity reflectivity $R=0.3$ \cite{Qudits}, as indicated in the Figure.}
  \label{fig:sfig2}
\end{subfigure}
\caption{Signatures in the HOM interference profile of the cavity induced time-frequency comb structure of the phase matching associated function of the generated photon pair.  }
\label{figComb}
\end{figure}

In first place, the facets of the non-linear waveguide are reflective, due to the refractive index contrast between semiconductor and air. For this reason, a cavity effect occurs, which produces a temporal grating that generates a frequency comb in the spectrum of the generated photons. The peaks of the comb are spaced by $\bar \tau /2$, where $\bar \tau$ is the time a photon takes to make a roundtrip in the cavity, and in \cite{Qudits} we have shown that by changing the pump frequency we can engineer different phase profiles for the comb, with different symmetry properties. An example can be seen in Fig. \ref{figComb}, that shows as well the dependency of the peak visibility with the reflectivity of the cavity, and the expected HOM profile for the existing device, for which the reflectivity $R \approx 0.3$.  This cavity effect enables for instance the discretization of such frequency combs, which are rising a considerable interest in the community, since they can be used  as qudits \cite{Qudits}, that can have applications in quantum key distribution, for instance \cite{Autebert_2016}.  

 % \begin{figure}
 %\begin{center}
%\includegraphics[width=0.5\textwidth]{FigComb.png}
%\caption{\label{figComb} }
%\end{center}
%\end{figure}

From the continuous variables perspective, large superpositions of numerous highly localized states can be used to encode error-protected qubits. Such states present redundancy and translational symmetry in the ideal case of an infinite superposition of infinitely localized states. These properties are interesting for quantum error correction, as first proposed by Gottesman-Kitaev and Preskill (GKP) \cite{PhysRevA.64.012310}. The so-called GKP states are robust against small displacements in phase space (small with respect to the comb spacing in position and momentum representation), and they can be used not only to encode quantum information but also as a ressource to correct displacement errors that affect different states defined in continuous variables \cite{PhysRevLett.123.200502}. In addition, it has been shown that it is a sufficient non-Gaussian element to complete Gaussian-based quantum computation and turn it universal \cite{PhysRevLett.123.200502}.

In spite of its several applications and fundamental interest for quantum computing with continuous variables, the experimental production of  GKP state is extremely challenging in the quadrature representation, since it involves creating highly non-Gaussian states. Nevertheless, some experimental results exist in the microwave range \cite{campagneibarcq:hal-03084673} and using the motional state of trapped ions \cite{IonGKP}. In \cite{PhysRevA.95.042311}, it was shown that spatial gratings can be used to engineer GKP states using the transverse position and momentum degrees of freedom of single photons. As for time and frequency variables, we can say for once that we are lucky: the cavity structure of AlGaAs semiconducting device naturally creates a comb structure, as mentioned. We have shown in \cite{PhysRevA.102.012607} that such comb can, indeed, be seen as an entangled GKP state, and measuring time or frequency of one the photons enables correcting the errors of the other, in a reminiscence of a measurement-based quantum computing scheme. 

The main idea behind our results is the fact that the effective non-linear interaction producing the photon pair in a cavity can be seen as a combination of universal gates, as defined in \cite{PhysRevA.105.052429}, acting on a separable ideal pair of GKP states. The effect of such gates is to add noise to the ideal state, under the form of a classical distribution of displacement operators, and entangle both states. Such operations can be seen as a small quantum circuit that produces entangled GKP states which were proven to be a resource for quantum error correction.

Additionally, frequency encoded GKP states can also be manipulated thanks to the HOM interferometer. As a matter of fact, in the subspace formed by GKP states, time and frequency displacements of fixed amounts, corresponding to the combs' interspacing, act as Pauli matrices do on qubits. Thus, by fixing the time delay in one of the arms of the HOM interferometer, one obtains the equivalent to the application of a quantum gate analogous to the Pauli matrix $\sigma_x$ to a GKP state, as was shown in \cite{PhysRevA.102.012607}.

Of course one can use different techniques to generate frequency comb structures, as placing a cavity in one arm of the HOM interferometer, as done in \cite{PhysRevLett.91.163602}. Nevertheless, using the natural cavity-like structure in the device studied in \cite{PhysRevA.102.012607} is interesting since it avoids extra sources of losses.

\section{Discussion and conclusion}

We have revisited the HOM experiment, a benchmark in quantum optics and quantum physics, taking a phase space perspective which enables interpreting time and frequency degrees of freedom of single photons as genuine quantum continuous variables, in perfect analogy with position and momentum. We also discussed the quantum nature of this experiment and how classical results reproducing the ones due to photon statistics may deserve to be interpreted under a different perspective. 

We have focused on a specific experimental set-up and shown how exotic states can be engineered in such a device by manipulating a classical pump beam. Also, we identified the ``natural" output of such a device to a entangled state with quantum properties which were shown to be useful for quantum error correction. The presented original approach to the HOM experiment opens the perspective to novel applications of single photon based protocols and shows that quantum phenomena haven't stopped surprising us. 

As an interesting perspective, we should mention the extension of the presented results to the atomic version of the HOM experiment \cite{AtomicHOM}, performed by Alain Aspect and co-workers.

\bibliography{biblio75}

\end{document}